\documentclass{article}

\usepackage{graphicx}		 
\usepackage{hyperref}
\usepackage[ruled,vlined]{algorithm2e}
\usepackage{listings}

\usepackage{authblk}

\usepackage[ngerman, british]{babel}

\begin{document}

\title{Toward Active and Passive Confidentiality Attacks On Cryptocurrency Off-Chain Networks}

\renewcommand*{\Authsep}{\qquad \qquad \qquad}
\renewcommand*{\Authand}{  }
\renewcommand*{\Authands}{\qquad \qquad \qquad \qquad}

\author[1]{Utz Nisslmueller}
\author[1]{Klaus-Tycho Foerster}
\author[1]{Stefan Schmid}
\author[2]{Christian Decker}

\renewcommand*{\Affilfont}{\normalsize}
\affil[1]{Faculty of Computer Science, University of Vienna, Vienna, Austria}
\affil[2]{Blockstream, Zurich, Switzerland}

\date{}
\maketitle

\begin{abstract} %%\noindent
Cryptocurrency off-chain networks such as Lightning (e.g., Bitcoin) or Raiden (e.g., Ethereum) aim to increase the scalability of traditional on-chain transactions. To support nodes in learning about possible paths to route their transactions, these networks need to provide gossip and probing mechanisms. This paper explores whether these mechanisms may be exploited to infer sensitive information about the flow of transactions, and eventually harm privacy. In particular, we identify two threats, related to an active and a passive adversary. The first is a \emph{probing attack:} here the adversary aims to detect the maximum amount which is transferable in a given direction over a target channel by actively probing it and differentiating the response messages it receives. The second is a \emph{timing attack:} the adversary discovers how close the destination of a routed payment actually is, by acting as a passive man-in-the middle and analyzing the time deltas between sent messages and their corresponding responses. We then analyze the limitations of these attacks and propose remediations for scenarios in which they are able to produce accurate results. 
\end{abstract}

\section{{Introduction}}
\label{sec:introduction}

Blockchains, the technology underlying cryptocurrencies such as Bitcoin or Ethereum, herald an era in which mistrusting entities can cooperate in the absence of a trusted third party. However, current blockchain technology faces a scalability challenge, supporting merely tens of transactions per second, compared to custodian payment systems which easily support thousands of transactions per second. This is the result of the underlying global consensus algorithms, which tread on the side of correctness rather than performance. \newline

%%\noindent
Off-chain networks~\cite{gudgeon2019sok}, a.k.a.\ payment channel networks (PCNs) or second-layer blockchain networks, have emerged as a promising solution to mitigate the blockchain scalability problem: 
by allowing participants to make payments directly through a network of \emph{peer-to-peer} payment channels, the overhead of global consensus protocols and committing transactions on-chain can be avoided.  Off-chain networks such as Bitcoin Lightning~\cite{lightningrfc}, Ethereum Raiden~\cite{raiden}, and XRP Ripple~\cite{fugger2004money}, to just name a few, promise to primarily reduce load on the underlying blockchain, as well as drastically increasing transaction throughput, and thus, being able to settle transactions in the matter of (sub-) seconds rather than in minutes or in hours - along with substantially reducing transaction fees, since now only one counterparty is responsible for validating a payment initially, rather than the whole network. \newline

%%\noindent
In all of these networks, each node typically represents a user and each weighted edge represents funds escrowed on a blockchain; these funds can be transacted only between the endpoints of the edge. Many payment channel networks use source routing, in which the source of a payment specifies the complete route for the payment. If the global view of all nodes is accurate, source routing is highly effective because it finds all paths between pairs of nodes. Naturally, nodes are likely to prefer paths with lower per-hop fees, and are only interested in paths which support their transaction, i.e. which have a sufficient channel capacity. \newline

%%\noindent
However, the fact that nodes need to be able to find routes also requires mechanisms for nodes to learn about the payment channel network's state. The two typical mechanisms which enable nodes to find and create such paths are \emph{gossip} and \emph{probing}. The gossip protocol defines messages which are to be broadcast in order for participants to be able to discover new nodes and channels and keep track of currently known nodes and channels \cite{bolt7}. \emph{Probing} is the mechanism which is used to construct an actual payment route based on a local network view delivered by gossip, and ultimately perform the payment. In the context of \S\ref{sec:probing_attack}, we are going to exploit probing to discover whether a payment has occurred over a target channel. The gossip store is queried for viable routes to the destination, based on the desired route properties \cite{getroute}. Because the gossip store contains global channel information, it is possible to query payment routes originating from any node on the network. Due to privacy concerns, gossip messages only include the \emph{total} balance for any given channel rather than the balance each node is holding. \newline

%%\noindent
This paper explores the question whether the inherent need for nodes to discover routes in general, and the gossip and probing mechanisms in particular, can be exploited to infer sensitive information about the off-chain network and its transactions.

\subsection{Our Contributions}

This paper identifies two novel threats for the confidentiality of off-chain networks. In particular, we consider the Lightning Network as a case study and present two attacks, an active one and a passive one. The active one is a \emph{probing attack} in which the adversary wants to determine the maximum amount which can be transferred over a target channel it is directly or indirectly connected to, by active probing. The passive one is a \emph{timing attack} in which the adversary discovers how close the destination of a routed payment actually is, by acting as a man-in-the middle and listening for / analyzing certain well-defined messages. We then analyze these attacks, identify limitations and also propose remediations for scenarios in which they are able to produce accurate results. 

\subsection{Organization}

Our paper is organized as follows. We introduce some preliminaries in \S\ref{sec:background},
and then first describe the probing attack in \S\ref{sec:probing_attack} followed by the timing attack in \S\ref{sec:timing_attack}. We review related work in \S\ref{sec:related_work} and conclude in \S\ref{sec:conclusion}. 

\section{Preliminaries}\label{sec:background}

While our contribution is applicable to the concept of off-chain networks in general, to be concrete, we will consider the Bitcoin Lightning Network (LN) as a case study in this paper. In the following, we will provide some specific preliminaries which are necessary to understand the remainder of this paper. \newline

%%\noindent
The messages which are passed from one Lightning node to another are specified in the Basics of Lightning Technology (BOLTs) \cite{bolts}. Each message is divided into a subcategory, called a layer. This provides superior separation of concerns, as each layer has a specific task and, similarly to the layers found in the Internet Protocol Suite, is agnostic to the other layers. \newline

%%\noindent
For example in Lightning, the \texttt{channel\textunderscore announce} and \texttt{channel\textunderscore update} messages are especially crucial for correct payment routing by other nodes on the network. \texttt{channel\textunderscore announce} signals the creation of a new channel between two LN nodes and is broadcast exactly once. \newline

%%\noindent
\texttt{channel\textunderscore update} is propagated at least once by each endpoint, since even initially each of them may have a different fee schedule and thus, routing capacity may differ depending on the direction the payment is taking (i.e., when $c$ is the newly created channel between A and B, whether $c$ is used in direction AB or BA). Once a viable route has been determined, the sending node needs to construct a message (a transaction ``request'') which needs to be sent to the first hop along the route. Each payment request is accompanied by an onion routing packet containing route information. Upon receiving a payment request each node strips one layer of encryption, extracting its routing information, and ultimately preparing the onion routing packet for the next node in the route. For the sake of simplicity, cryptographic aspects are going to be omitted for the rest of this chapter. We refer to \cite{bolt4} and \cite{bolt8} for specifics. \newline

\begin{figure}[h]
\centering
\includegraphics[width=\textwidth]{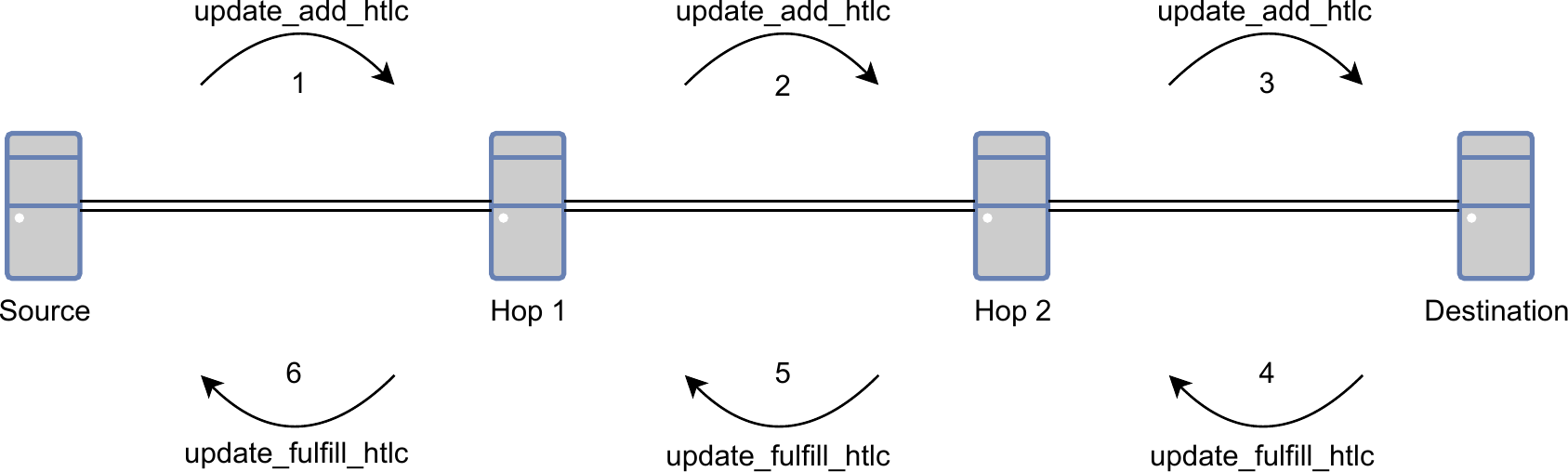}
\caption{An exemplary transaction from source to destination, involving two intermediate nodes.}
\label{fig:add_fulfill_htlc}
\end{figure}

\noindent
Two BOLT Layer 2 messages are essential in order to to establish a payment chain:
\begin{itemize}
  \item \texttt{update\textunderscore add\textunderscore htlc}: This message signals to the receiver, that the sender would like to establish a new HTLC (Hash Time Locked Contract), containing a certain amount of millisatoshis, over a given channel. The message also contains an \texttt{onion\textunderscore routing\textunderscore packet} field, which contains information to be forwarded to the next hop along the route. In Figure~\ref{fig:add_fulfill_htlc}, the sender initially sets up an HTLC with Hop~1. The \texttt{onion\textunderscore routing\textunderscore field} contains another \texttt{update\textunderscore add\textunderscore htlc} (set up between Hop 1 and Hop 2), which in turn contains the ultimate \texttt{update\textunderscore add\textunderscore htlc} (set up between Hop 2 and Destination) in the \texttt{onion\textunderscore routing\textunderscore field}.
  
  \item \texttt{update\textunderscore fulfill\textunderscore htlc:} Once the payment message has reached the destination node, it needs to release the payment hash preimage in order to claim the funds which have been locked in the HTLCs along the route by the forwarded \texttt{update\textunderscore add\textunderscore htlc} messages. For further information on why this is necessary and how HTLCs ensure trustless payment chains, see \cite{mastering_bitcoin}. To achieve this, the preimage is passed along the route backwards, thereby resolving the HTLCs and committing the transfer of funds (see Steps 4, 5, 6 in Figure~\ref{fig:add_fulfill_htlc}).
\end{itemize}

%\noindent
The gossip messages mentioned earlier are sent to every adjacent node and eventually propagate through the entire network. 

\texttt{update\textunderscore add\textunderscore htlc} and \texttt{update\textunderscore fulfill\textunderscore htlc} however, are only sent/forwarded to the node on the other end of the~HTLC. \newline

\begin{figure}[h]
\centering
\includegraphics[width=8cm]{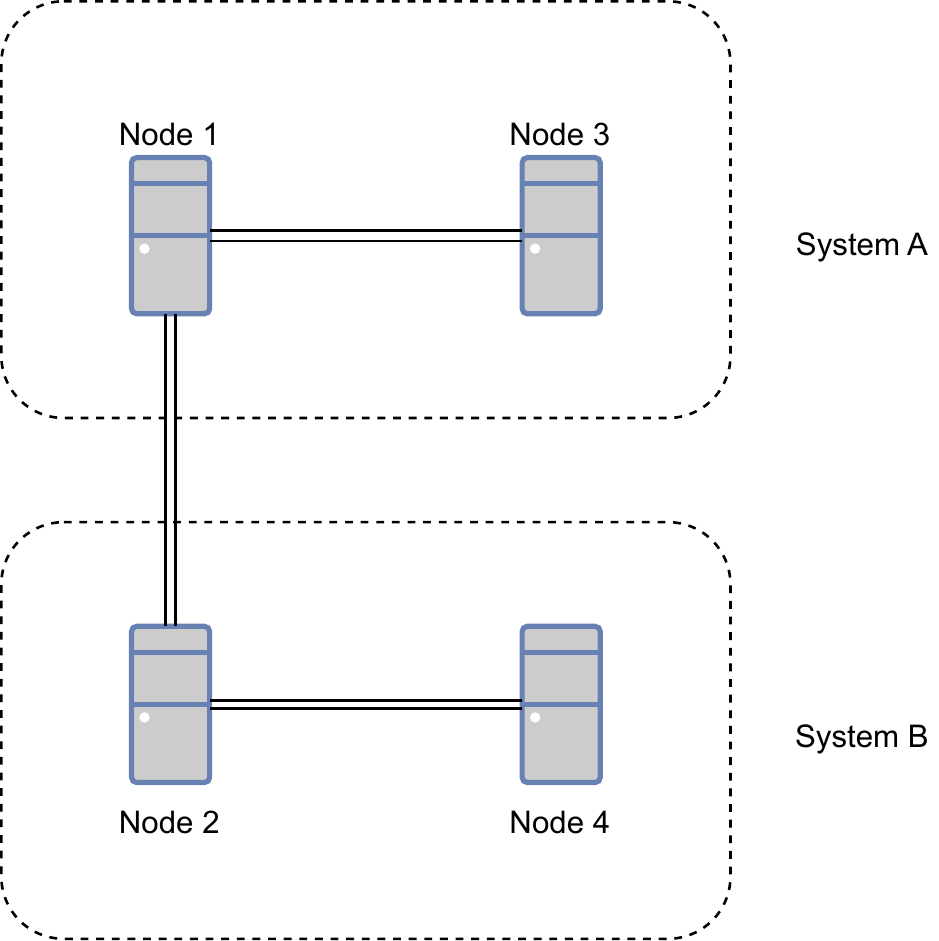}
\caption{Local Testing Setup}
\label{fig:setup}
\end{figure}

%\noindent
In order to test the attacks proposed in  \S\ref{sec:probing_attack} and \S\ref{sec:timing_attack}, we have set up a testing network consisting of four c-lightning \cite{clightninggit} nodes, with two local network computers running two local nodes each (Figure \ref{fig:setup}). Nodes 1 and 2 are connected via a local network link and can form hops for payment routes between Nodes 3 and 4. In order to interact with the nodes, we have made use of c-lightning's RPC interface and built our software tool set in Python \cite{python_github}. The tests and their corresponding results in §\ref{sec:timing_attack} have also been verified with LND \cite{lndgit}, another BOLT-conform Lightning Network implementation, written in Go.

\section{Related Work}
\label{sec:related_work}

Off-chain networks in general and the Lightning network in particular have recently received much attention, and we refer the reader to the excellent survey by Gudgeon et al.~\cite{gudgeon2019sok}. The Lightning Network as an second-layer network alternative to pure on-chain transactions was first proposed by \cite{ln_whitepaper}, with the technical specifications laid out in \cite{lightningrfc}. Despite being theoretically currency-agnostic, current implementations such as c-lightning \cite{clightninggit} and LND \cite{lndgit} support BTC exclusively. A popular alternative for ERC-20 based tokens is the Raiden Network \cite{raiden}. \newline

%\noindent
Several papers have already analyzed security and privacy concerns in off-chain networks. Rohrer et al.~\cite{rohrer2019discharged} focus on channel-based attacks and proposes methods to exhaust a victim's channels via malicious routing (up to potentially total isolation from the victim's neighbors) and to deny service to a victim via malicious HTLC construction. Tochner et al.~\cite{tochner2019hijacking} propose a denial of service attack by creating low-fee channels to other nodes, which are then naturally used to route payments for fee-minimizing network participants and then dropping the payment packets, therefore forcing the sender to await the expiration of the already set-up HTLCs. \newline

%\noindent
\cite{balancehiding} provides a closer look into the privacy-performance trade-off inherent in LN routing. The authors also propose an attack to discover channel balances within the network. Wang et al. \cite{wang2019flash} examine the LN routing process in more detail and proposes a split routing approach, dividing payments into large size and small size transactions. The authors show that by routing large payments dynamically to avoid superfluous fees and by routing small payments via a lookup mechanism to reduce excessive probing, the overall success rate can be maintained while significantly reducing performance overhead. Beres et al. \cite{beres2019cryptoeconomic} make a a case for most LN transactions not being truly private, since their analysis has found that most payments occur via single-hop paths. As a remediation, the authors propose partial route obfuscation/extension by adding multiple low-fee hops. Currently still work in progress, \cite{lnbook} is very close to \cite{mastering_bitcoin} in its approach and already provides some insights into second-layer payments, invoices and payment channels in general. The Lightning Network uses the Sphinx protocol to implement onion routing, as specified in \cite{bolt4}. The version used in current Lightning versions is based on \cite{sphinx1} and \cite{sphinx2}, the latter of which also provides performance comparisons between competing protocols.

\section{Probing Attack}
\label{sec:probing_attack}

\subsection{Design}
\label{ssec:probing_design}

The Lightning Network uses an invoice system to handle payments. A LN invoice consists of a destination node ID, a label, a creation timestamp, an expiry timestamp, a CLTV (Check Lock Time Verify) expiry timestamp and a payment hash. Paying an invoice with a randomized payment hash is possible (since the routing nodes are yet oblivious to the actual hash) and will route the payment successfully to its' destination, which forms the basis of this attack. Optionally it can contain an amount (leaving this field empty would be equal in principle to a blank cheque), a verbal description, a BTC fallback address in case the payment is unsuccessful, and a payment route suggestion. This invoice is then encoded, signed by the payee, and finally sent to the~payer. \newline

%\noindent
Having received a valid invoice (e.g.\ through their browser or directly via e-mail), the payer can now either use the route suggestion within the invoice or query the network themselves, and then send the payment to the payee along the route which has been determined. In this section, we will use the c-lightning RPC interface via Python exclusively - the functions involved are \texttt{ getroute()}~\cite{getroute} and \texttt{sendpay()}~\cite{sendpay}, which takes two arguments: the return object from a \texttt{getroute()} call for a given route, a given amount and a given riskfactor, as well as the payment hash. Using \texttt{sendpay()} on its own (meaning, with a random payment hash instead of data from a corresponding invoice) will naturally result in one of two following error codes:

\begin{itemize}
    \item \textbf{204 (failure along route):} This error indicates that one of the hops was unable to forward the payment to the next hop. This can be either due to insufficient funds or a non-existent connection between two adjacent hops along the specified route. If we have ensured that all nodes are connected as depicted in Figure~\ref{fig:setup}, we can safely assume the former. One sequence of events leading up to this error can be seen in Figure~\ref{fig:error204}.

\begin{figure}[h]
\centering
\includegraphics[width=\textwidth]{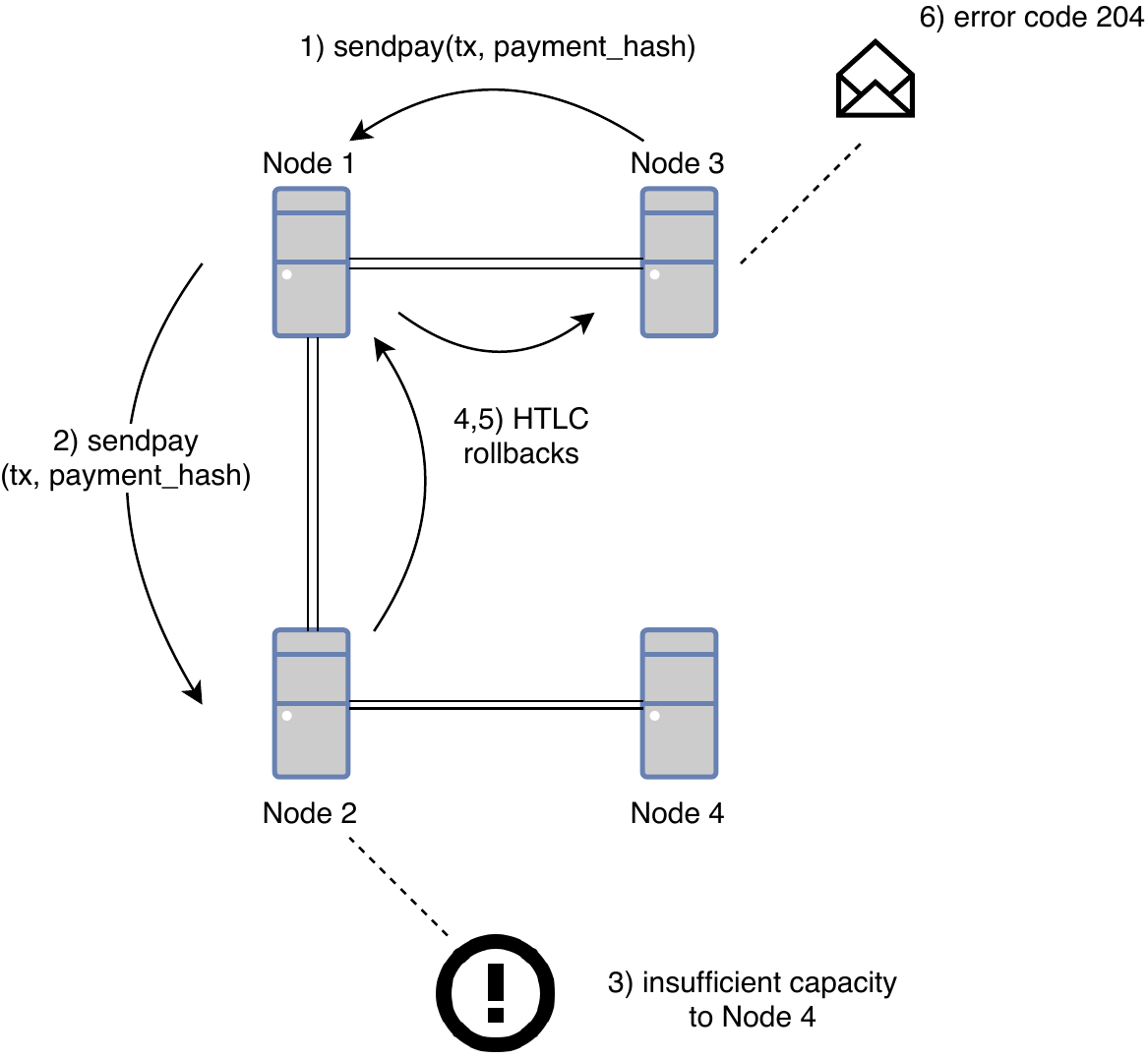}
  \caption{Causing a 204 error by trying to send a payment to Node 4, which Node 3 is unable to perform.}
  \label{fig:error204}
\end{figure}
    
    \item \textbf{16399 (permanent failure at destination):} Given the absence of a 204 error, the attempted payment has reached the last hop. As we are using a random payment hash, realistically the destination node will throw an error, signalling that no matching preimage has been found to produce the payment hash. The procedure to provoke a 16399 error code can be seen in Figure~\ref{fig:error16399}.
    
\begin{figure}[h]
\centering
\includegraphics[width=\textwidth]{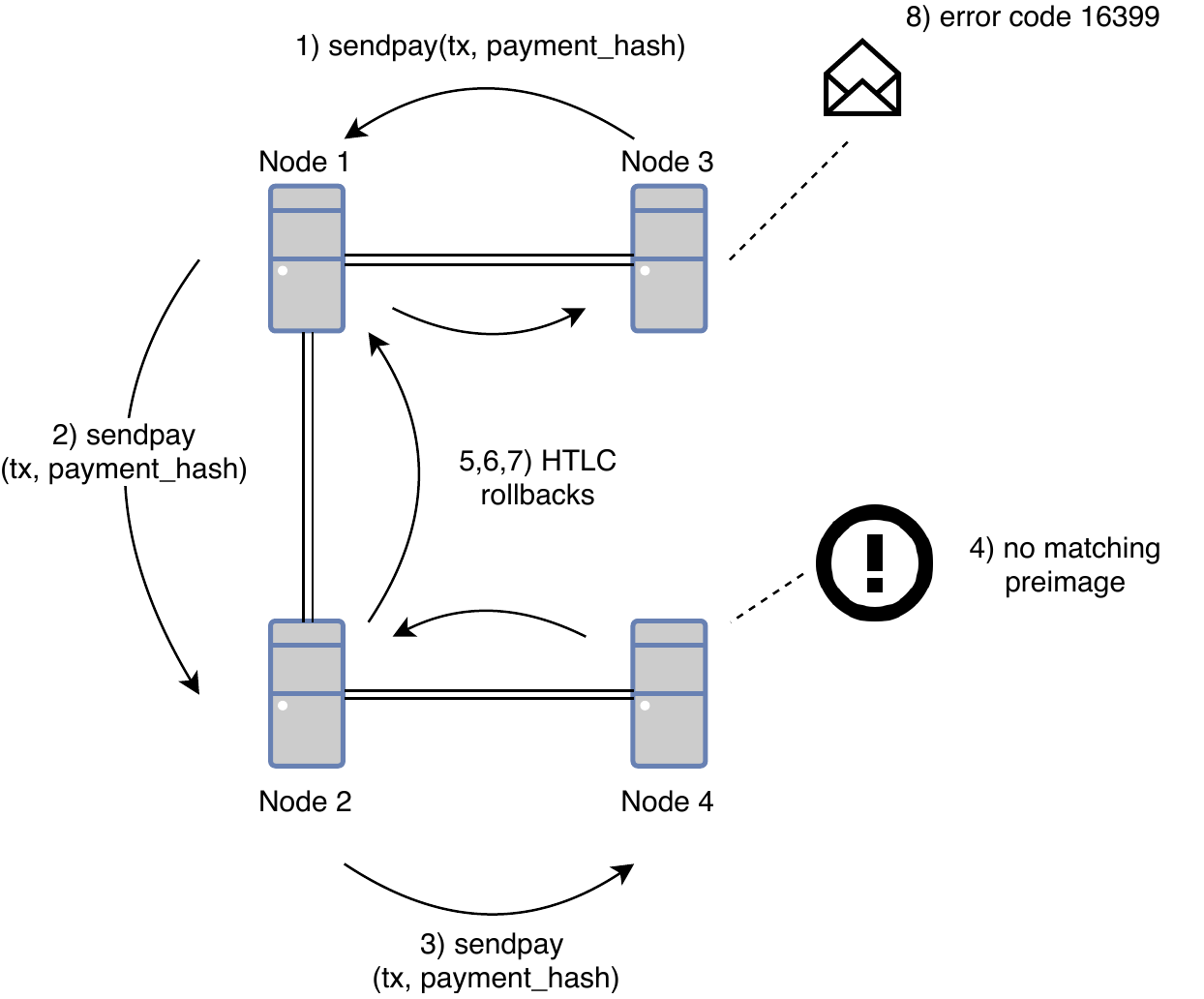}
    \caption{Causing a 16399 error by trying to send a payment to Node 4, who can't produce a matching preimage and thus fails the payment.}
    \label{fig:error16399}
\end{figure}
    
\end{itemize}

%\noindent
The goal of this attack is to trace payment flow over a channel, which the attacker node is directly or indirectly connected to. The attacker node  will therefore initially attempt to determine whether a payment has occurred over the observed channel between the penultimate and final node along the route. To this end, the attacker will send out periodic probes to the final node (the "victim"), containing the amount which has been determined by the initial probe. If channel weights remain unaltered, each of these probes should return a 16399 error code. If a payment does occur however, the penultimate node will find itself unable to forward the payment on the outgoing channel to our target, yielding a 204 error response. Upon receiving this message, we can then restart the process of our initial probe and ultimately arrive at the exact amount of millisatoshis (msat), which have been transferred. 

\subsection{Lab Implementation}
\label{ssec:probing_lab_implementation}

Recalling Figure \ref{fig:setup}, we have chosen Node 3 as our attacker node and Node 4 as our target node - hence, the initial goal of Node 3 is to determine the maximum payment flow between Nodes 2 and 4. To conduct our tests, each of the channels has been set up with a balance of 200,000,000 msat, with each node holding a stake of 100,000,000 msat in each of its channels. Node 3 will hold a slightly higher balance in order to accommodate probing fees. We can use the total channel balance, as received via gossip, as an upper ceiling for this value (200,000,000 msat in this case). We can then send payments from Node 3 to Node 4 with random payment hashes - resulting in either error code 16399 or error code 204 (\S \ref{ssec:probing_design}). To this end, we perform a binary search on the available funds which we can transfer, searching for the highest value yielding a 16399 error instead of a 204 error. The algorithms used for both initial probing and deriving the actual channel balance from Node 2 to Node 4 are depicted in Algorithms \ref{alg:probe} and \ref{alg:find_init_max}.\newline

\begin{algorithm}[H]
\SetAlgoLined
\KwResult{Either error code 204 or 16399}
 payment\textunderscore hash = random.hex()\;
 node\textunderscore id = node ID of final node on victim channel\;
 msat = \textit{value to probe for}\;
 route = getroute(node\textunderscore id, msat)\;
 sendpay(route, payment\textunderscore hash)\;
 \caption{Probing a channel for a given amount of msat}
 \label{alg:probe}
\end{algorithm}\

\begin{algorithm}[H]
\SetAlgoLined
\KwResult{amount\textunderscore msat - initial channel balance}
 min\textunderscore msat = 0\;
 max\textunderscore msat = channel.balance\;
 amount\textunderscore msat = channel.balance / 2\;
 \While{True}{
 \
  \eIf{probe(amount\textunderscore msat) == 16399}{
   min\textunderscore msat = amount\textunderscore msat\;
   }{
   \eIf{amount\textunderscore msat) == 204}{
   max\textunderscore msat = amount\textunderscore msat\;
   }{
   return "No suitable route found."\;
   }
  }
  \eIf{max\textunderscore msat - min\textunderscore msat < 1000}{
   return amount\textunderscore msat\;
   }{
   // \textit{continue to minimise maximum error}
   }
  amount\textunderscore msat = (min\textunderscore msat + max\textunderscore msat) / 2
 }
 \caption{Finding the initial maximum channel balance}
 \label{alg:find_init_max}
\end{algorithm}

\vspace{10px}
%\noindent
We thus arrive at the approximate maximum amount, which Node 2 can transfer to Node 4. The next step is to continuously probe for this amount of msat in regular intervals. The expected response is a 16399 error code, with a 204 error code implying that the amount we are trying to send is higher than the available amount which Node 2 can transfer to Node 4 (or that it has disconnected from Node 4). Upon receiving a 204 response, we start looking for the maximum payable amount to Node 4 once more. Subtracting the new amount from the old amount, we arrive at the size of the transaction which has occurred between Nodes 3 and 4. \newline

%\noindent
After 17 probes by Node 3, Algorithm \ref{alg:find_init_max} has yielded an initial balance of 99,999,237 msat, which is in line with the channel balance we have allocated between Nodes 2 and 4. The next step is to monitor the channel for potential weight changes (Algorithm \ref{alg:monitor balance}). \vspace{10px}

\begin{algorithm}[H]
\SetAlgoLined
\KwResult{New maximum flow from penultimate to final node}
 init\textunderscore max = \textit{initial channel balance}\;
 new\textunderscore max = init\textunderscore max\;
 t = \textit{time to wait between checks}\;
 \While{True}{
sleep(t)\;
  \eIf{probe(init\textunderscore max) == 204}{
  // \textit{channel balance has decreased} \\
   return find\textunderscore init\textunderscore max()\;
   // \textit{potentially calculate delta} \\
   }{
   \eIf{(init\textunderscore max + 1000) == 16399}{
   // \textit{channel balance has increased} \\
   return find\textunderscore init\textunderscore max()\;
   // \textit{potentially calculate delta} \\
   }{
   return error\;
   }
  }
 }
 \caption{Finding the initial maximum channel balance}
 \label{alg:monitor balance}
\end{algorithm}

\vspace{10px}
%\noindent
To verify this, we have transferred 50,000,000 msat from Node 2 to Node 4, with our program detecting this soon after (we have set t to 5 seconds in order to avoid excessive probing) and returning an updated balance of 49,998,237 msat. We then transferred another 30,000,000 msat from Node 1 to Node 4, with our program again picking up the change and reporting the new channel balance at 19,997,389 msat. \newline

\begin{figure}[h]
\centering
\includegraphics[width=\textwidth]{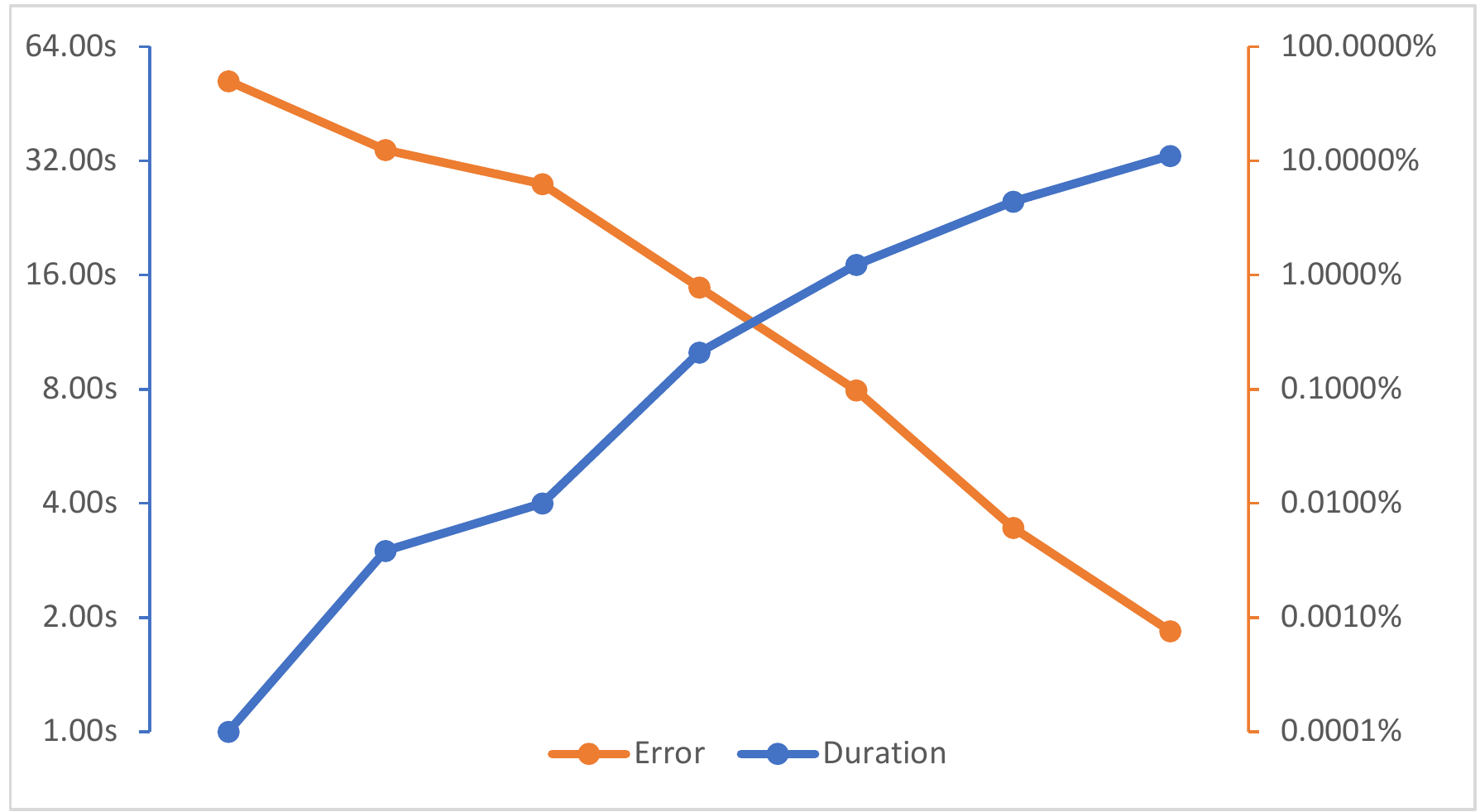}
  \caption{Visualizing the trade-off between probing accuracy and duration.}
  \label{fig:probing_graph}
\end{figure}

%\noindent
Figure \ref{fig:probing_graph} shows the trade-off between probing run time and the error in the channel balance estimate we observed for test runs on our lab setup. As we wanted to avoid overly excessive probing while conducting our tests, we were generally satisfied with any answer which is less than 1000 msat (the actual minimum BTC denomination) lower than the actual channel balance. Another possible approach could be keeping the number of probes sent out to the target constant, hence providing a more uniform level of balance error and probing duration. \newline

\subsection{BTC Testnet Evaluation}
\label{ssec:probing_testnet_evaluation}

\begin{figure}[h]
\centering
\includegraphics[width=\textwidth]{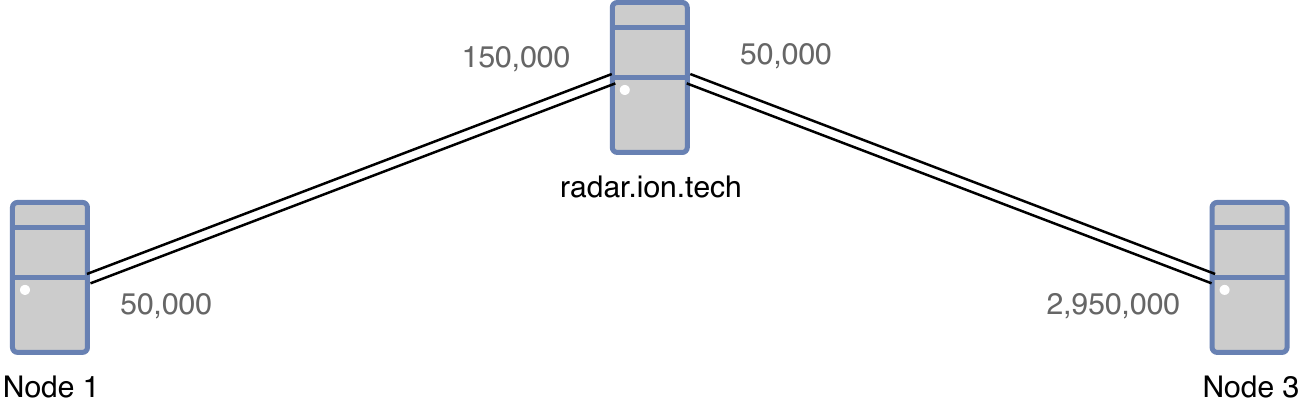}
  \caption{Setup and balance allocations of our first testnet evaluation (balances given in satoshis).}
  \label{fig:probing_setup_1}
\end{figure}

For analysis on the feasibility of our attack over the BTC Testnet, we connected Nodes 1, 2 and 3 from Figure \ref{fig:setup} to the "ion.radar.tech" Testnet Lightning node. We chose this host in particular, since their website allowed us to alter the channel weights by generating payable invoices with parameters of our choosing. The exact connections along with the corresponding channel weights can be seen in Figure \ref{fig:probing_setup_1}. Our goal was to verify the results we obtained in §\ref{ssec:probing_lab_implementation} and see whether probing duration (see Figure \ref{fig:probing_graph}) was affected by the public Testnet hop in place of the local hop(s) used in §\ref{ssec:probing_lab_implementation}. Running an initial series of probes from Node 3 to Node 1, we arrived at a channel balance of 149,926,757 msat between the radar.ion.tech node and Node 1 (99.95\% accuracy). We attribute this comparatively high error in regard to our tests in §\ref{ssec:probing_lab_implementation} due to the Testnet nodes' differing fee structure, which is necessarily taken into account when constructing the payment route. Then, we sent a payment containing 50,000,000 msat from Node 2 to Node 1 - predictably, Node 3 returned the updated maximum payment flow on the observed channel correctly with 99,902,343 msat (99.9 \% accuracy). \newline

\begin{figure}[h]
\centering
\includegraphics[width=\textwidth]{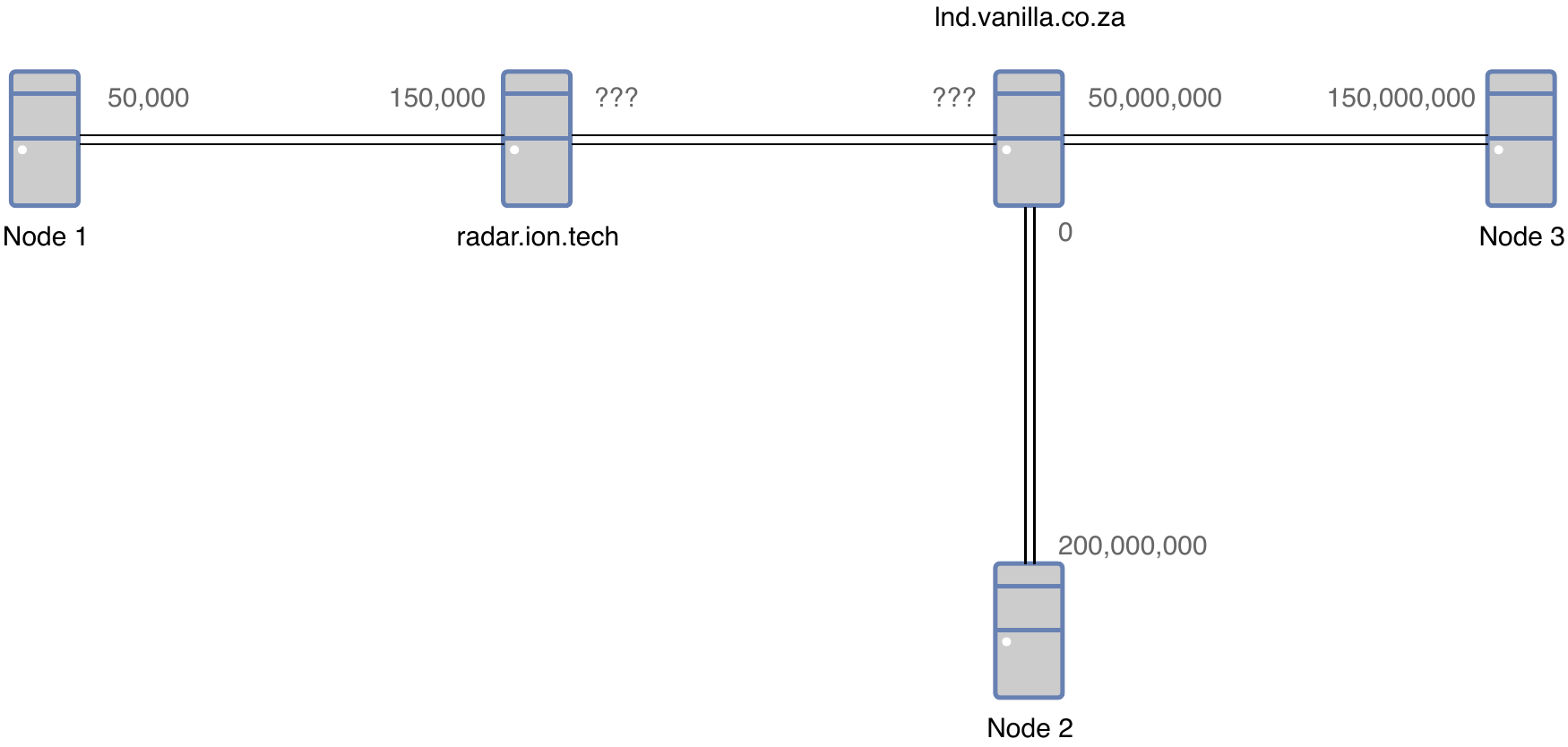}
  \caption{Initial setup and balance allocations of our second Testnet evaluation (balances given in satoshis).}
  \label{fig:probing_setup_2a}
\end{figure}

%\noindent
After verifying the correct operation of our program for 16399 error codes, we were keen on discovering whether 204 error code scenarios would be dealt with correctly as well. In order to test this, we transferred back any amounts which have been redistributed as part of our initial test, increased the channel balance between Node 1 and radar.ion.tech by a factor of 10 and modified the setup from Figure \ref{fig:probing_setup_1} slightly by placing an intermediary hop between radar.ion.tech and Nodes 2 and 3. The updated infrastructure can be seen in Figure \ref{fig:probing_setup_2a}. \newline

%\noindent
It became apparent however, that we would need to rethink the weights we allotted to the respective nodes, as we were initially unaware of the true channel weights between the radar.ion.tech and "lnd.vanilla.co.za" nodes. Naturally, we were inclined to simply run the \texttt{find\textunderscore init\textunderscore max()} function (Algorithm \ref{alg:find_init_max}) from Node 3 on the ion.radar.tech node. However, we found that the two nodes were connected by 6 channels rather than one. To circumvent this route ambiguity, we queried a route for 1,000, 1,000,000 an 1,000,000,000 msat using default parameters, hoping all of them would return the same route, thus allowing us to treat the resulting channel as the only one connecting these two nodes. Unfortunately though, we received varying responses for all of these amounts, introducing  a large uncertainty in any subsequent measurements. We then tried to run our tests on these channels, with all of them reporting failure in establishing a route to the target. We are not sure why even the initial probes failed and only further analysis and testing of our program will unveil the error in our approach. We decided to conclude our Testnet evaluation at this point, since despite extensive refactoring, we were not able to produce meaningful results for this constellation of nodes and channels, leaving route ambiguity and handling of multiple channels to be explored by further research in this area.

\subsection{Results, Implications, and further Considerations}
\label{ssec:probing_results}

In  \S\ref{ssec:probing_lab_implementation} we have demonstrated that it is in fact possible to trace channel payments if the network is structured in a certain way. In theory, this method should hold true for any node which is reachable from the attacking node and has only one channel whose balance is lower or or equal to the second lowest balance on the route from the attacking node. We have partially verified this supposition in §\ref{ssec:probing_testnet_evaluation} while maintaining a high accuracy in our successful measurements. This is particularly a threat to end users, since most of them connect to a single well-connected node over a single channel, in order to interact with the rest of the network \cite{1ML}. Nonetheless, there are several caveats to this method, the most significant of which are:

\begin{itemize}
    \item \textbf{Excluding the possibility of payment forwarding:} The attack laid out in this chapter does not take into account the fact that nodes can be used to forward payments. Hearkening back to Figure \ref{fig:setup}, if we were to select the channel between Nodes 1 and 2 as our target, transactions between Nodes 1 and 4 would appear as if they were transactions to Node 3. One opportunity of accounting for this would be to monitor \emph{every} channel to and from Node 3 for changes in directed channel balance, which would create problems on its own (see below).
    
    \item \textbf{Surge of unresolved HTLCs while probing:} Recalling steps 5-7 in Figure \ref{fig:error16399}, each probe sets up a chain of irredeemable HTLCs (since a matching preimage would have to be brute-forced). Eventually, running multiple probes over the same channels will escrow its funds in these HTLCs, effectively DOSing the probe route and forcing the nodes to wait until the HTLCs time out before being able to forward other payments. This is an issue we encountered over and over during §\ref{ssec:probing_lab_implementation} and §\ref{ssec:probing_testnet_evaluation}, often giving us one shot at probing before having to wait multiple hours for the HTLCs to expire. This is also why we chose the channels leading up to our final target to have a much higher balance, so that we would have enough balance left after initial probing to monitor the channel for a reasonable period of time.

    \item \textbf{Insufficient sensitivity for high-frequency transactions:} Looking back at Algorithm \ref{alg:monitor balance}, we have defined the parameter t as the time, for which to wait during probes for monitoring the channel balance, one the initial maximum value has been discovered. If more than one transaction would occur during this timeframe, it would still only show up as a singular payment with our tool. In the worst-case scenario, two transactions covering the same amount could take place in opposite directions, not changing the weighted balance at all and thus eluding our detection mechanisms.

    \item \textbf{Omission of private channels:} Upon creating a channel, the node can declare the channel as private, and thus prevent it from being broadcast via gossip. The channel is fully functional for both nodes which are connected by it, but no foreign payments can be routed through it. Looking ahead to increasing adoption of the Lightning Network, this provides an intriguing opportunity for nodes, which do not wish to participate in routing (e.g.\ mobile wallets) or nodes with limited uptime (personal computers). Routing would only occur between aggregating nodes (such as payment providers), with most of the channels (and therefore nodes) on the network remaining invisible to malicious participants as the gossip protocol would only propagate public channels. This further exacerbates our ability to detect forwarded payments (see above) as opposed to actual payments, since private channels can't be monitored by design.

    \item \textbf{Disregard of potential bottlenecks:} The proposed method of monitoring channel transactions does not hold, if a single channel along the route has a lower balance than the target channel in the desired direction. The node which has an insufficient amount of msat on its' outgoing channel would return a 204 error (Figure \ref{fig:error204}) This can often happen if an end user node is used as a hop prior to a high-capacity node. It is easy to detect which channel acts as a bottleneck, however a bit trickier to circumvent this obstacle - we would like to point the interested reader to \cite{tochner2019hijacking} for suggestions on route hijacking and thus effectively bypassing the bottleneck along the route.
\end{itemize}

%\noindent
During the tests we conducted in §\ref{ssec:probing_testnet_evaluation}, we also encountered the hops between Node 1 and Node 3 being connected via multiple channels. As confirmed by our observations, it is entirely possible to receive varying routes for differing amount\textunderscore msat, riskfactor, cltv and fuzzpercent \cite{getroute} combinations. Our tool failed to produce accurate results in this scenario, as it was designed assuming singular channels between pairs of nodes. It is however perfectly reasonable to have multiple channels between two nodes, as channel balances are final and can't be increased after creation. We expect this to be the predominant form of retrospectively increasing potential payment flow between nodes and further research on how to deal with this complication would be highly appreciated. \newline

%\noindent
All in all, the probing attack we laid out in this chapter can be seen more as a proof of concept rather than a realistic attack vector, due to the limitations discussed in §\ref{ssec:probing_results}. We are confident that certain aspects such as the exact algorithm and route construction could be refined to provide more reliable results. However other aspects such as the binding of channel funds in irredeemable HTLCs and the incomplete network view due to private channels provide a much more consistent barrier to uncovering payment flows in real-world scenarios.

\section{Timing Attack}
\label{sec:timing_attack}

\subsection{Design}
\label{ssec:timing_design}
\begin{figure}[h]
\centering
\includegraphics[width=\textwidth]{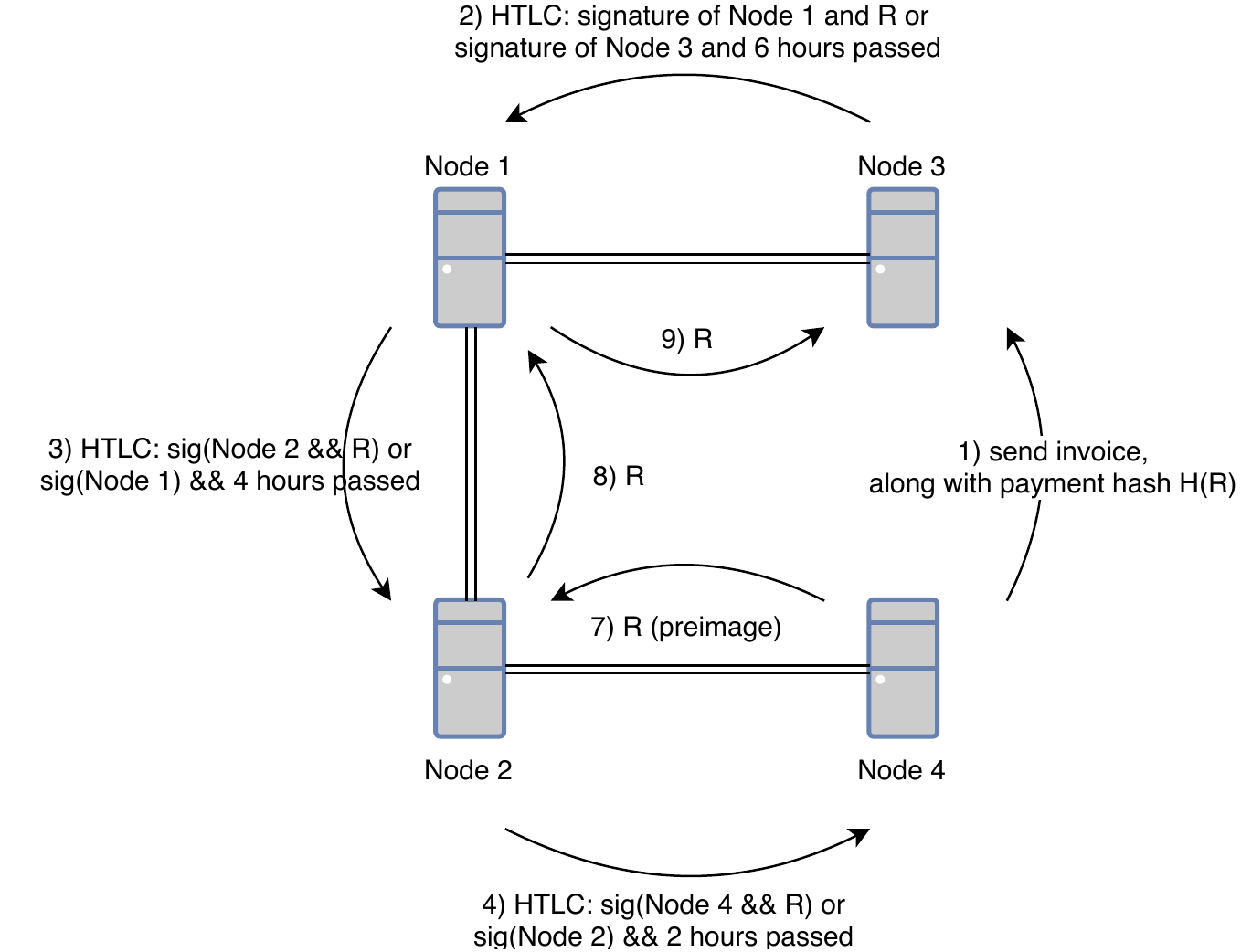}
  \caption{Paying a LN invoice over multiple hops. Messages 2-4 are \texttt{update\textunderscore add\textunderscore htlc} messages, messages 7-9 are \texttt{update\textunderscore fulfill\textunderscore htlc} messages. }
     \label{fig:htlc_chain}
\end{figure}

The Lightning Network is often referred to as a payment channel network (PCN). Performing payments over multiple hops is possible due to the use of HTLC's~\cite{ln_whitepaper}, a special bitcoin transaction whose unlocking conditions effectively rid the Lightning Network and its users of all trust requirements. An exemplary chain of HTLCs along with their shortened unlocking conditions is shown in Figure \ref{fig:htlc_chain}. Note that any node can only retrieve the funds locked in the HTLCs if they share R, and that each HTLC starting from Node 4 is valid for 2 hours longer than the previous HTLC to provide some room for error/downtime. \newline

%\noindent
Due to the Onion Routing properties of the Lightning Network, it is cryptographically infeasible to try and determine where along the route a forwarding node is located, since each node can only decrypt the layer which was intended for it to decrypt. Attempts to analyze the remaining length of the routing packet have been thwarted at the protocol level by implementing a fixed packet size with zero padding at the final layer~\cite{bolt4}. \newline

%\noindent
The only opportunity left to analyze the encrypted traffic between the nodes is to extract time-related information from the messages. One possibility would be to analyze the \texttt{cltv\textunderscore expiry\textunderscore delta} field (analogous to ``hours passed'' in Figure \ref{fig:htlc_chain}, measured in mined blocks since the establishment of the HTLC): By looking at the delay of both the incoming and the outgoing HTLC, a node could infer how many hops are left until the payment destination. However, this possibility has been accounted for by the adding "shadow routes" to the actual payment path, with each node fuzzing path information by adding a random offset to the \texttt{cltv\textunderscore expiry\textunderscore delta} value, hence effectively preventing nodes from guessing their position along the payment route \cite{bolt7}. \newline

%\noindent
The method we propose, is to time messages at the network level, rather than at the protocol level (e.g. through \texttt{cltv\textunderscore expiry\textunderscore delta}). Recalling Figure \ref{fig:htlc_chain}, Node 2 can listen for response messages from Node 4, since there is currently no mechanism in place to add delay to \texttt{update\textunderscore fulfill\textunderscore htlc} responses (in fact, \cite{bolt2} states that ``\textit{a node SHOULD remove an HTLC as soon as it can}''). Based on response latency, Node 2 could infer its position along the payment route to a certain extent, as examined in \S\ref{ssec:timing_implementation}.

\subsection{Lab Implementation}
\label{ssec:timing_implementation}
Initial analysis has shown that analyzing packets directly (e.g.\ via Wireshark) is of little avail, since LN messages are end-to-end encrypted - meaning that even if we know the target nodes' IP address and port number, we can not detect the exact nature of the messages exchanged. We hence chose to redirect the output of the listening c-lightning node to a log file, which we then analyze with a Python script. As in \S\ref{sec:probing_attack}, the source code can be found at \cite{python_github}. \newline

% cdecker: Falls gewuenscht kann ich einen htlc_resolved hook einbauen der getriggert wird wenn ein eingehender HTLC von uns gesettled wird. Zusammen mit dem bereits existierenden htlc_accepted hook kann man damit das genaue Timing eines HTLCs errechnen ohne logs parsen zu muessen.
% Kommentar: Das wäre für Future Work, oder?

\begin{figure}[h]
\centering
\includegraphics[width=\textwidth]{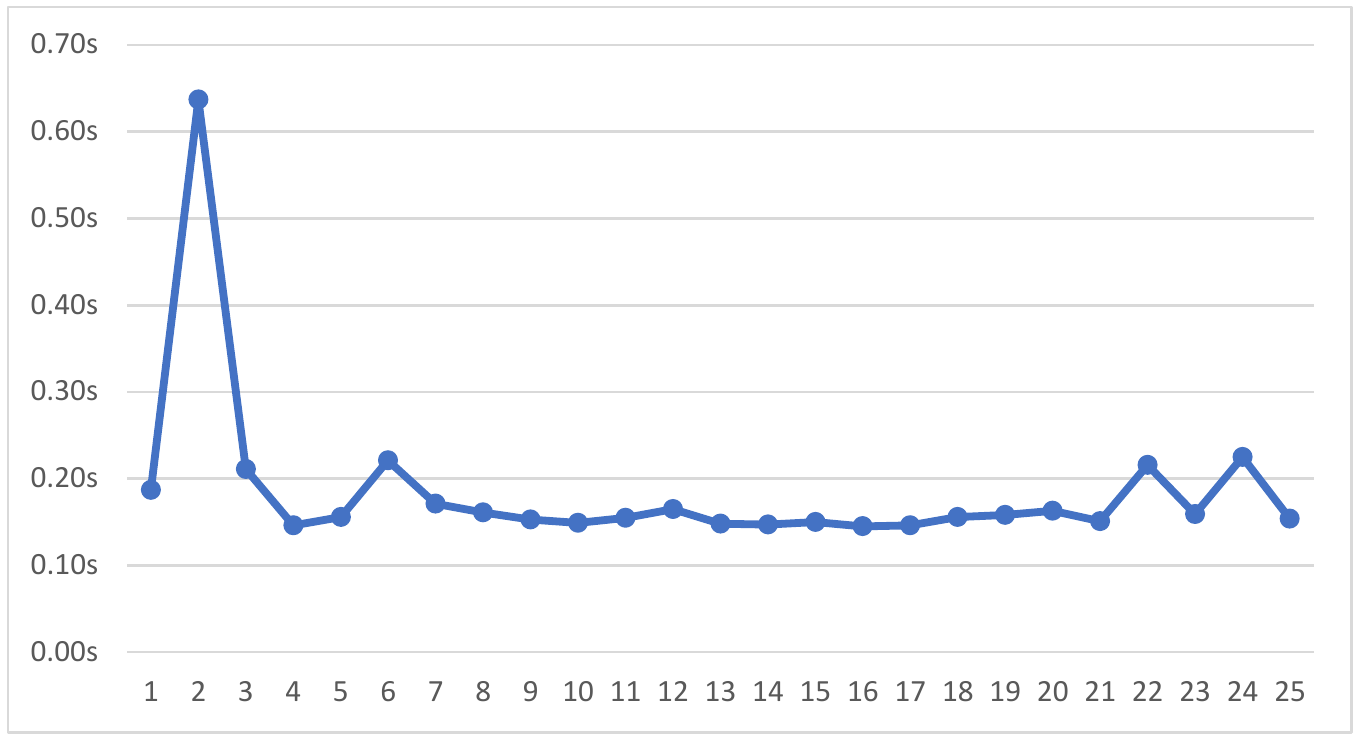}
  \caption{Latency times for local payments containing 1,000 msat ($\mu$ = 0.1852, $\sigma$ = 0.0974, $n$ = 25)}
     \label{fig:small_payments}
\end{figure}

\begin{figure}[h]
\centering
\includegraphics[width=\textwidth]{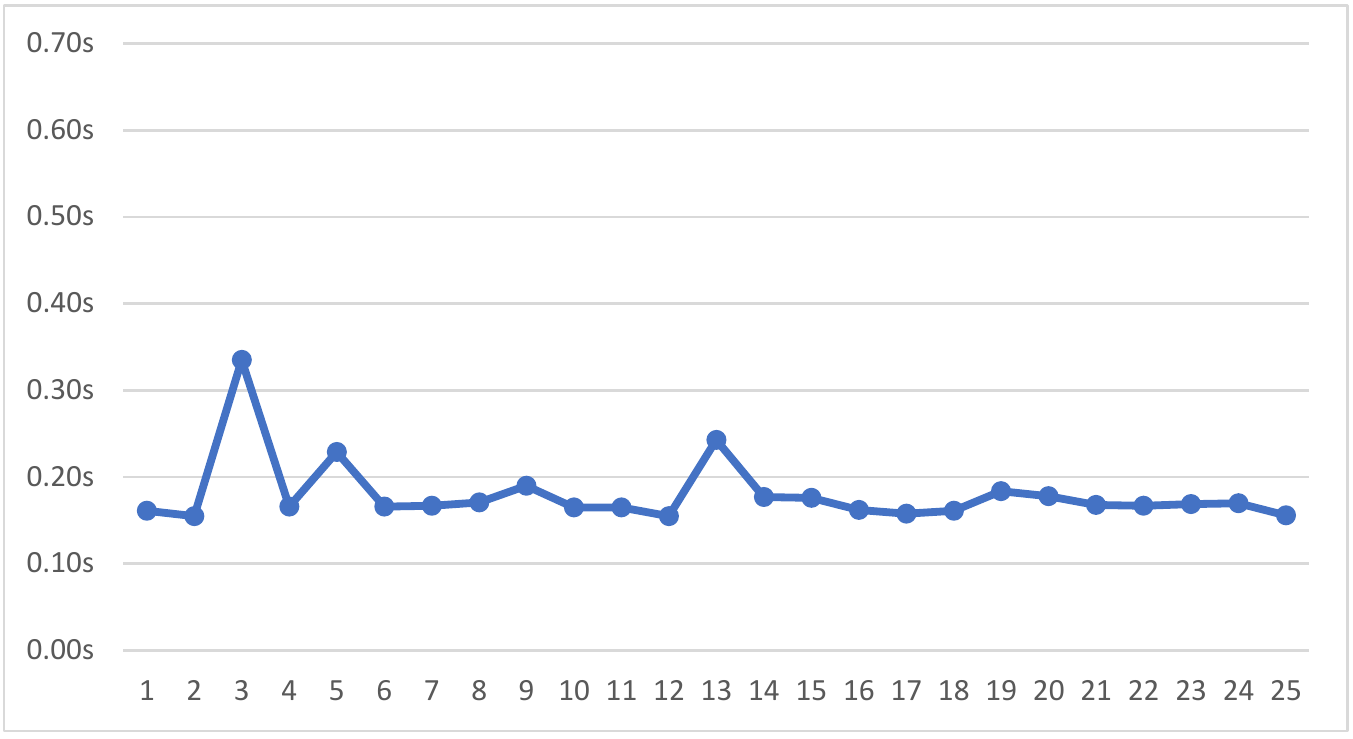}
   \caption{Latency times for local payments containing 100,000,000 msat ($\mu$ = 0.1798, $\sigma$ = 0.0385, $n$ = 25)}
     \label{fig:large_payments}
\end{figure}

%\noindent
Looking at the log file, we are particularly interested in the two messages discussed in \S\ref{sec:background}: \texttt{update\textunderscore add\textunderscore htlc} and  \texttt{update\textunderscore fulfill\textunderscore htlc}. The node output includes these events, complete with timestamps and the corresponding node ID with which the HTLC is negotiated. By repeatedly sending money back and forth between Nodes 1 and 3 in our test setup (Figure \ref{fig:setup}), we arrive at a local (and therefore minimum) latency of 182ms on average. The latency distribution for small (1,000 msat) payments can be seen in Figure~\ref{fig:small_payments}. We have found that latencies remain largely unaffected by transaction size - increasing payment size by a factor of 100,000 actually slightly reduced average settlement time and standard deviation~(Figure~\ref{fig:large_payments}). \newline

\begin{figure}[!h]
\centering
\includegraphics[width=\textwidth]{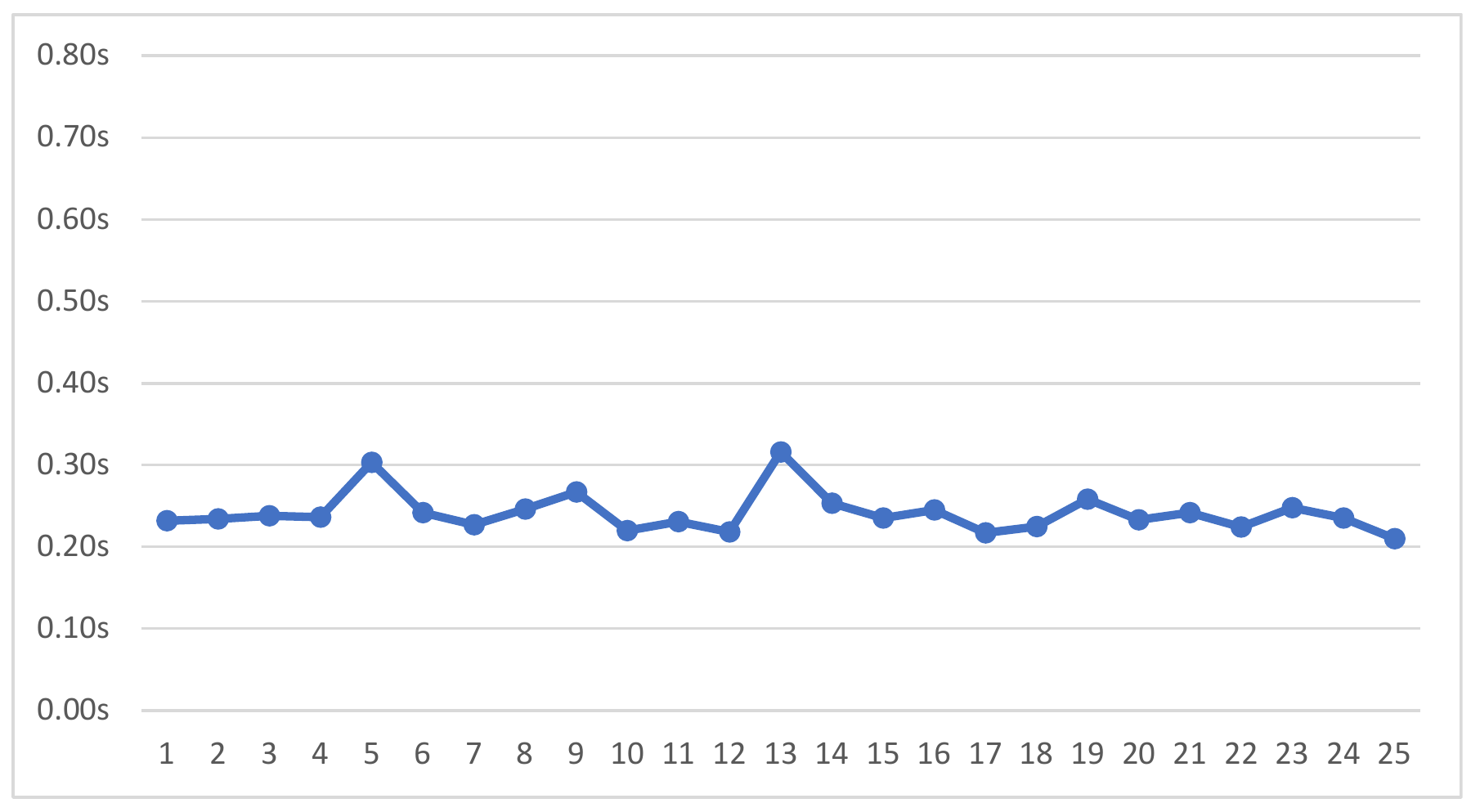}
   \caption{Latency times for payments containing 100,000,000 msat over 1 network hop ($\mu$ = 0.234, $\sigma$ = 0.025, $n$ = 25)}
     \label{fig:node2node1}
\end{figure}

\begin{figure}[!h]
\centering
\includegraphics[width=\textwidth]{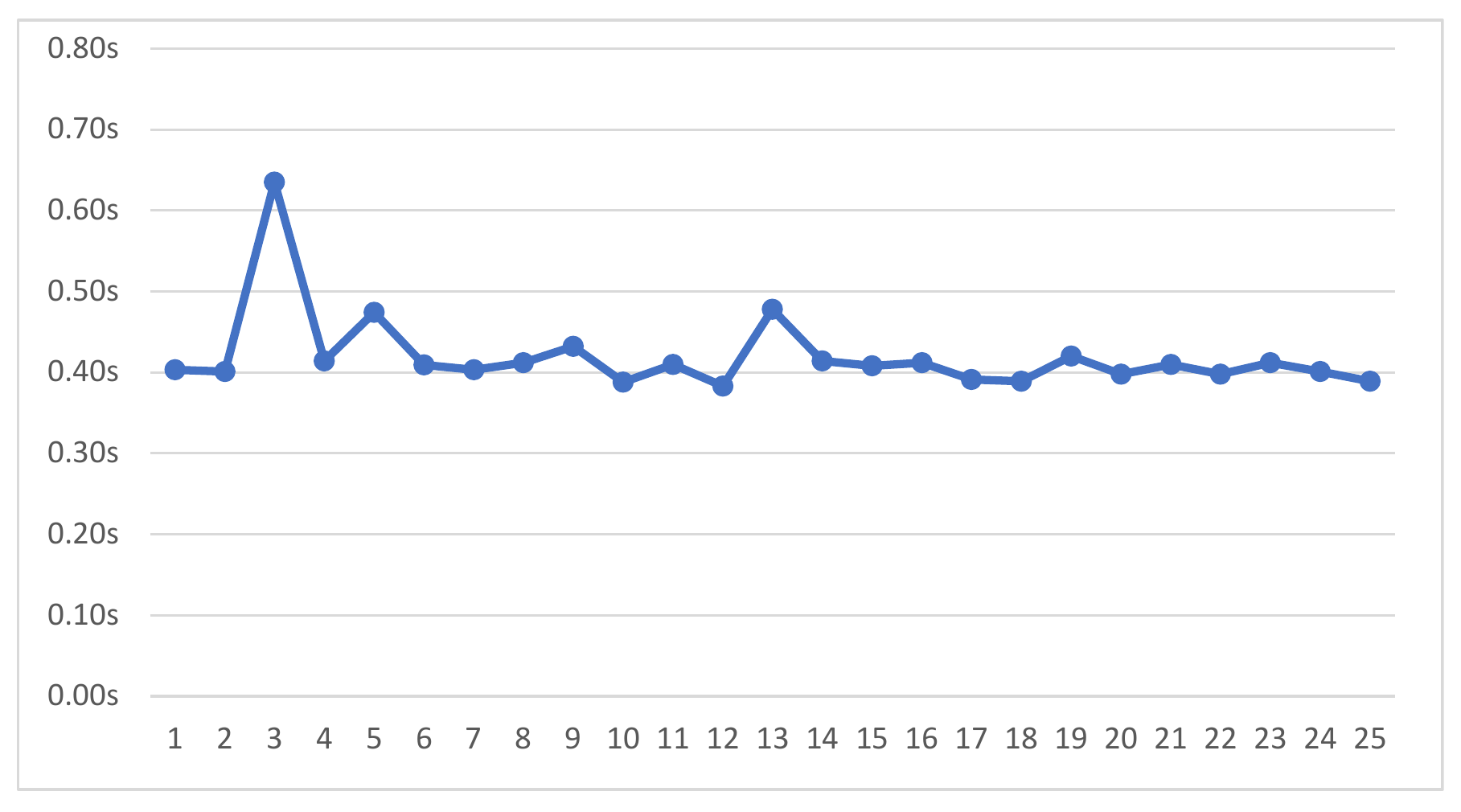}
   \caption{Latency times for payments containing 100,000,000 msat over 1 network hop and 1 local hop ($\mu$ = 0.414, $\sigma$ = 0.05, $n$ = 25)}
     \label{fig:node2node3}
\end{figure}

%\noindent
Next, we examined whether an increase in hop distance would yield predictable results. To this end, we first timed payments over 1 network hop from Node 2 to Node 1 (Figure \ref{fig:node2node1}). Then, we timed payments over the same amount over 1 network and 1 local hop from Node 2 to Node 3 (Figure \ref{fig:node2node3}). Based on these results, we derive that timing messages on a local network with little to no interfering traffic scales predictably over several hops, with 1 network hop roughly corresponding to 1.284 local hops in terms of latency. 

\subsection{BTC Testnet Evaluation}
\label{ssec:timing_testnet_evaluation}
Building on the results obtained in §\ref{ssec:timing_implementation}, we were keen to discover whether the they would carry over into real-world evaluations. To this end, we connected Node 1 and Node 3 from Figure \ref{fig:setup} to the "endurance" Lightning Testnet node. Located in Dublin, Ireland and being connected to over 500 other Lightning Testnet nodes \cite{1ML}, we concluded that this node would provide a good entry point to test network latency from our location in Vienna, Austria, with the possibility to construct longer and more complicated routes over it as we saw fit. In order to constitute an initial RTT value, we established an HTTP connection to Lightning's default port 9735 \cite{bolt1}, since the target host appeared to drop our ICMP ping requests. Alternating our requests between Systems A and B (Figure \ref{fig:setup}) in an attempt to prevent cached responses, we have found that HTTP response times were fairly constant from this node, with an average response time of 0.067s ($\sigma$ = 0.0206).\newline

\begin{figure}[!h]
\centering
\includegraphics[width=\textwidth]{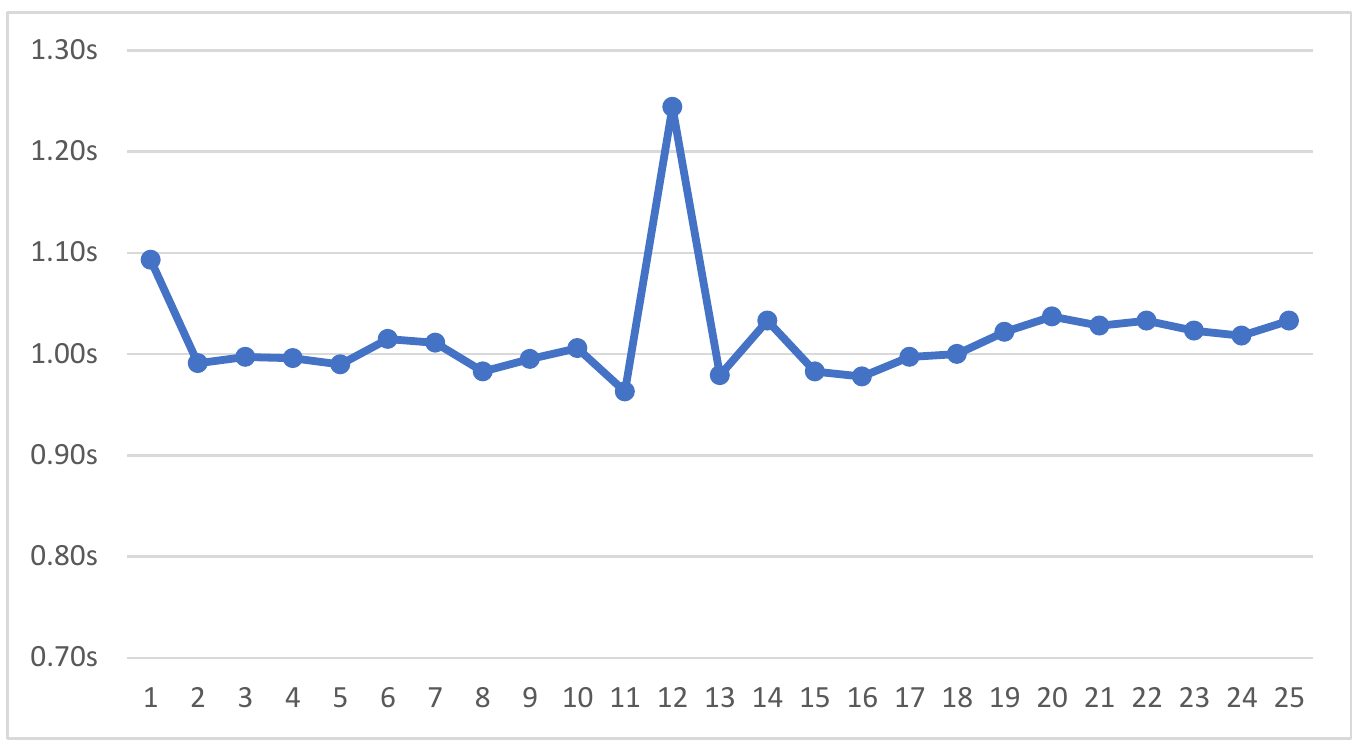}
   \caption{Latency times for payments between Node 1 and Node 3 over the endurance Lightning node ($\mu$ = 1.0179, $\sigma$ = 0.0542, $n$ = 25)}
     \label{fig:endurance_latency}
\end{figure}

%\noindent
Next, we were interested whether payments over the public hop were subject to an equally uniform latency as in §\ref{ssec:timing_implementation}. Thus, we created 25 invoices over 1,000,000 msat each (having found in §\ref{ssec:timing_implementation} that response latency is independent of payment size) at Node 3 and sent the payment from Node 1. As seen in Figure \ref{fig:endurance_latency}, the fulfill message response times were remarkably consistent, however latency did not scale to our expectations. Based on Figure \ref{fig:node2node3}, we expected to be overall latency to be in the ballpark of 0.5-0.7 seconds (2x local network RTT + HTTP request RTT), however actual latency was twice that value. Results from the aranguren.org Testnet node, located in Melbourne, Australia, proved equally consistent with an average ping time of 0.314s ($\sigma$ = 0.035) and an average HTLC fulfillment latency of 1.68s ($\sigma$ = 0.0972) \newline % maybe ask why latency was higher than expected?

\begin{figure}[!h]
\centering
\includegraphics[width=\textwidth]{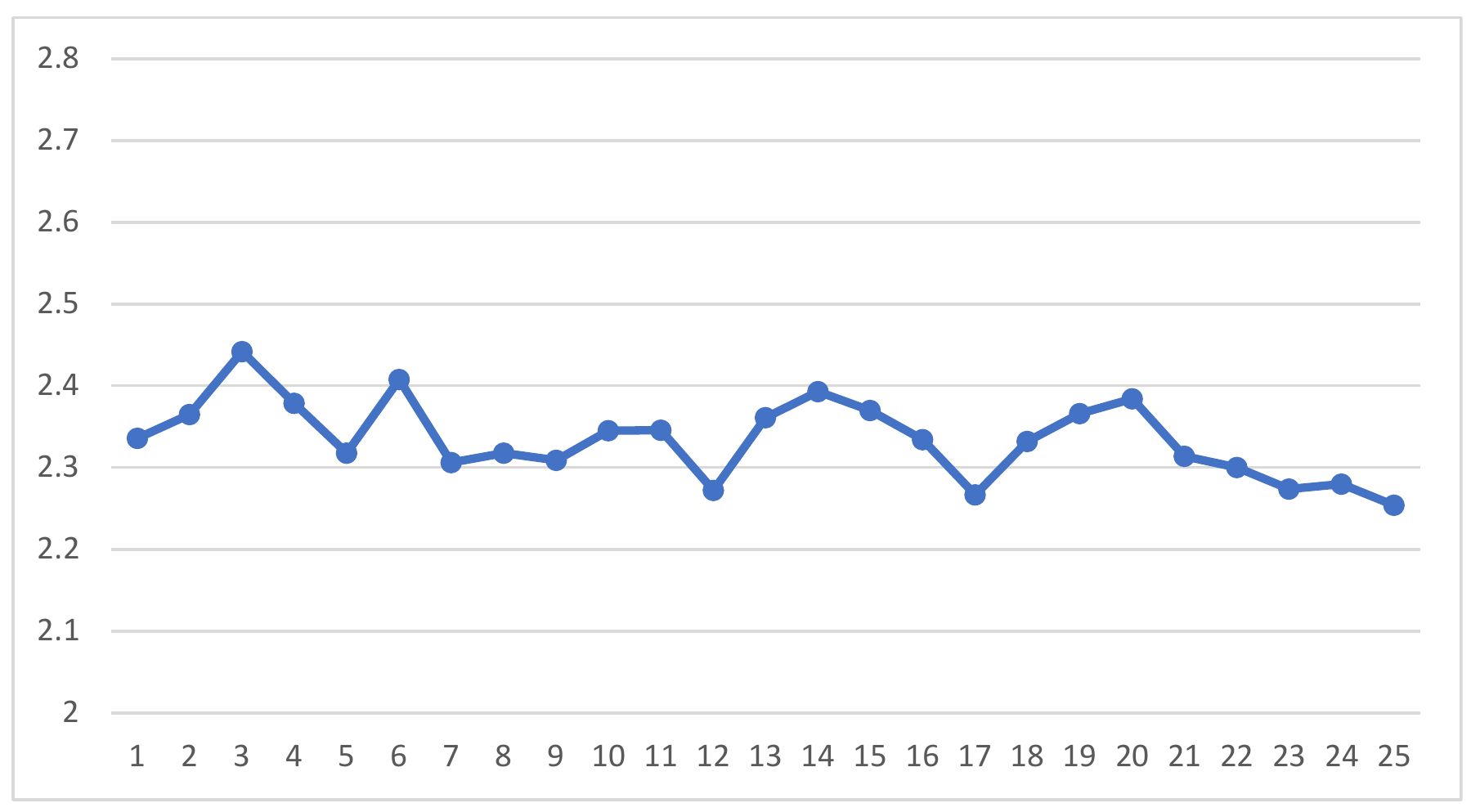}
   \caption{Latency times for payments between Node 1 and Node 3 over the endurance and aranguren.org Lightning nodes ($\mu$ = 2.3349, $\sigma$ = 0.0475, $n$ = 25)}
     \label{fig:2_hop_latency}
\end{figure}

%\noindent
Finally, we were curious about HTLC fulfillment delays over 2 public hops. To this end, we closed the channel between Node 3 and endurance and opened a new channel to the "aranguren.org" Testnet node, which in turn has a channel with endurance and thus re-establishes the chain of channels from Node 1 to Node 3. Timing results for this route can be seen in Figure \ref{fig:2_hop_latency}. This marked the end of our timing tests, since we were not able to establish an acyclic payment route over 3 or more publicly available LN nodes. This coincides with the observation that neither the attempted nor the actual payments we performed during the course of §\ref{sec:probing_attack} and §\ref{sec:timing_attack} were routed over more than two public hops.

\subsection{Results, Implications and further Considerations}
\label{ssec:timing_results}
Considering the findings in \S\ref{ssec:timing_implementation}, we can see that timing produces fairly reliable and uniformly distributed results over a local network with little outside interference. Yet, due to the nature of LN routing, it is not possible to determine the distance or path to the initial payment source. To our surprise however, RTT remained equally consistent over 1-2 internet hops. Data acquired during monitoring of the local (mostly idle) network suggests that the timing node won't be able to distinguish traffic originating from a local node from the traffic in \S\ref{ssec:timing_implementation} without further information due to low latency deltas ranging from 2ms to 5ms. \newline 

%\noindent
While performing timing measurements for payments across the BTC Testnet network, we have found that HTLC settlement takes long enough over even 1 hop to make traffic RTT volatility negligible. Over 1 hop, we conclude that HTLC settlement for our Vienna-based node should be in the ballpark of 0.86 - 1.97 seconds with 2-hop latency amounting to roughly 1.99 - 2.68 seconds, depending on the geographical location and assuming a normal distribution for the measured latency deltas. Further research could include a further statistical examination of the ability to differentiate distances for HTLC deltas at the sub-2-second threshold. We suggest that overall network bandwidth does not affect the acquired results significantly, since after performing all payments in §\ref{sec:probing_attack}, Node 1 has sent 64 KB and Node 3 has received 55 KB - only a fraction of which were outgoing/incoming HTLCs (alongside gossip, pings, etc.). \newline

%\noindent
Our results open many new avenues for further timing-based research on the Lightning Network. The next step for us would be to develop a tool to predict the distance to the final destination of an HTLC which is passing through the listening node, based on the measurements laid out in §\ref{ssec:timing_testnet_evaluation}. It would be interesting to see whether there is a possibility to force payment-unrelated response messages, e.g. by forging ping messages \cite{bolt1} in order to estimate (possibly network-wide) RTTs, correlate HTLC settlement latencies against them and finally arrive at a set of nodes which must have been the ultimate recipient of the forwarded payment. Furthermore, experiments could be conducted on the feasibility of adding a random time offset to HTLC fulfillments, and the trade-offs involved therein.

% explanation: 0.855-1.97 -> endurance and aranguren 1 hop range +- 3SD

\section{Conclusion}
\label{sec:conclusion}

This paper has shown that off-chain routing and payment settlement mechanisms may be exploited to infer confidential information about the network state. In particular, considering the Lightning Network with Bitcoin as the underlying blockchain as a case study, we set up a local infrastructure and proposed two ways in which two current state-of-the-art implementations, c-lightning and LND, can be exploited to gain knowledge about distant channel balances and transactions to unconnected nodes: By deliberately failing payment attempts, we were able to deduce the exact amount of (milli-)satoshis on a channel located two hops away on our local lab infrastructure. Using this technique repeatedly, we were able to determine whether a transaction occurred between one node and another over the monitored channel. To a certain extent, we were able to reproduce these results in the public Bitcoin Testnet chain. We also identified this attacks' limitations and proposed some workarounds to these obstacles. \newline

%\noindent
By timing the messages related to HTLC construction and termination, we were able to infer the remaining distance of a forwarded packet accurately in our test lab. These results transferred well into our Testnet evaluation, while being free of the partially restrictive limitations which we discovered during our examination of the probing attack. We concluded that RTT volatility of the HTLC message cycle was low enough for public Testnet hops which were within geographical vicinity to our node in Vienna, Austria, as well as for hops which were located in East Asia, to establish latency approximate latency boundaries for the number of remaining hops along the payment route of a forwarded transaction. \newline

%%\noindent
Our work raises several interesting research questions. In particular, it remains to fine-tune our attacks, to improve the flexibility of our software tools and to finally conduct more systematic experiments including more natural/interconnected network topologies, particularly on other off-chain networks. More generally, it will be interesting to explore further attacks on the confidentiality of off-chain networks exploiting the routing mechanism and investigate countermeasures. Furthermore, our work raises the question whether such vulnerabilities are an inherent price of efficient off-chain routing or if there exist rigorous solutions. 

\medskip

\noindent\textbf{Bibliographical Note.}
A preliminary version of this article appears at ICISSP 2020~\cite{icissp20}.
The work herein is based on the thesis of Utz Nisslmueller~\cite{thesis}.

%\noindent

%\newpage

\bibliographystyle{acm}
\bibliography{bibliography}

\end{document}